\documentclass[aps,pra,epsfigure,twocolumn,notitlepage,longbibliography]{revtex4-1}
\usepackage{dcolumn}    
\usepackage{bm} 
\usepackage{graphicx}
\usepackage{amsmath}    
\usepackage{latexsym}
\usepackage{amsfonts}   
\usepackage{amssymb}
\usepackage{array}      
\usepackage{epsfig}
\usepackage{txfonts}
\usepackage{color}
\usepackage[colorlinks=true,linkcolor=blue,urlcolor=blue,citecolor=blue]{hyperref}
\usepackage{hyperref}

\newcommand{\ket}[1]{\left\vert#1\right\rangle}
\newcommand{\bra}[1]{\left\langle#1\right\vert}

\begin{document}
\title{Ergotropy from coherences in an open quantum system}

\author{Bar\i\c{s} \c{C}akmak}
\email{baris.cakmak@eng.bau.edu.tr}
\email{cakmakb@gmail.com}
\affiliation{College of Engineering and Natural Sciences, Bah\c{c}e\c{s}ehir University, Be\c{s}ikta\c{s}, Istanbul 34353, Turkey}
\date{\today}
\begin{abstract}
We show that it is possible to have non-zero ergotropy in the steady-states of an open quantum system consisting of qubits that are collectively coupled to a thermal bath at a finite temperature. The dynamics of our model leads the qubits into a steady-state that has coherences in the energy eigenbasis when the system consists of more than a single qubit. We observe that even though the system does not have inverted populations, it is possible to extract work from the coherences and we analytically show that in the high temperature limit, ergotropy per unit energy is equal to the $l_1$ norm of coherence for the two qubit case. Further, we analyze the scaling of coherence and ergotropy as a function of the number of qubits in the system for different initial states. Our results demonstrate that one can design a quantum battery that is charged by a dissipative thermal bath in the weak coupling regime.
\end{abstract}
\maketitle
\section{Introduction}

The field of quantum thermodynamics generalizes the definitions of quantities such as heat, work and entropy that are made for macroscopic systems, to the realm of microscopic quantum systems and their dynamics~\cite{DeffnerCampbellBook,JPA_Goold,Alicki2018,CP_Vinjanampathy}. In other words, it is a field that analyzes the thermodynamics of non-equilibrium quantum processes. 

One of the most striking features of quantum systems is the fact that they can be in coherent superpositions of their possible available states. Despite its significance for quantum systems, a robust scheme to quantify coherence has only been recently introduced~\cite{PRL_Baumgratz} and attracted considerable interest~\cite{RMP_coherence}. Building on this scheme many works have demonstrated that quantum coherence can be used as an advantage or a resource for various thermodynamic processes~\cite{Science_scully,PRL_Aaberg,PRA_Uzdin,Korzekwa2016,PRX_Llobet,francica,francica2017,Entropy_Mauro,Entropy_ceren,PRA_Angsar,JPA_deVincente,NJP_Streltsov,PRA_Rana,PRE_Brandner,PRE_Angsar,DublinCoherence,arXiv_Alan,PRL_Kwon} (see also~\cite{RPP_Lostaglio,RMP_coherence}). Nevertheless, the inevitable interaction of a quantum system with its environment results in the loss of such genuine quantum features~\cite{RMP_Zurek}, together with the advantages they bring, therefore limiting their utilization in real physical settings. The dynamics of such systems are treated in the well-established formalism of open quantum systems~\cite{BreuerPetruccione,AlickiLendiBook}.

Within the realm of quantum thermodynamics one topic that attracts a significant attention is quantum batteries~\cite{Campaioli2018}. A quantum battery is a quantum system that a finite amount of its energy can be extracted as work through unitary cyclic processes~\cite{EPL_Allahverdyan,AlickiFannes}. There are mainly two sides to the literature in quantum batteries: charging and work extraction~\cite{AlickiFannes,PRL_Hovhannisyan,JPB_Steve,NJP_Binder,PRL_Campiaoli,PRL_Ferraro,PRA_Le,Quantum_Friis,PRE_Zhang,PRB_Andolina,PRL_Andolina,PRB_Rossini,PRB_Andolina2,PRR_Caravelli,PRR_Gherardini,PRA_Ghosh,PRX_Llobet,francica,francica2017,arXiv_Rossini,arXiv_Rosa,EPL_Dou,EPL_Allahverdyan2}. 
In both of these processes the main goal is to design or find dynamics that the energy is imparted or extracted from the battery in a feasible way according to certain figures of merit such as the total energy involved, power, etc. While it is naturally the case for work extraction processes, the vast majority of works examining charging protocols also rely on unitary evolution. Recently, however, a number of works have explicitly taken into account environmental interactions~\cite{PRB_Farina,PRE_Alan,PRA_Molmer,arXiv_Kamin,PRE_Alan_Sarandy,PRL_Barra,arXiv_Hovhannisyan,arXiv_Quach,arXiv_Alan}. Among the literature on quantum batteries, \cite{PRL_Barra,arXiv_Hovhannisyan} and \cite{PRX_Llobet,Korzekwa2016,francica,francica2017,DublinCoherence,arXiv_Alan,PRL_Kwon} particularly stand out for their relation to the present work, since they consider charging processes that only involve interaction with an environment and analyze the effects of coherences and/or correlations on the amount of extractable work, respectively.

In this work, we consider a collection of two-level systems (qubits) that are collectively interacting with a thermal environment. It has been shown that the steady-state of such a system can maintain coherences in its energy eigenbasis, if the number of qubits is greater than one~\cite{PRA_Angsar,PRA_Latune}, due to indistinguishably of the constituents of the system to the bath in the collective coupling regime. We show that the presence of these coherences results in a finite ergotropy of these steady-states starting from two qubit systems. We present analytical results on the coherence and ergotropy for an arbitrary number of qubits when they are initiated in their ground state, and numerically analyze the behavior for random initial states up to seven qubits. Our findings present a robust way to generate and store ergotropy in a quantum battery through coherences in an open system setting by purely thermal means while assuming a weak, but collective, coupling between the battery and its environment. 

The remainder of the paper is organized as follows. In Sec.~\ref{prelim} we introduce the figures of merit that we will be mainly interested in throughout this work. Sec.~\ref{model} describes the dynamical model which generates the coherences and the steady-states with finite ergotropy that we will use in our discussion. We examine the ergotropy of the steady-states of our model for two qubits and how it compares with the coherence at different temperatures and initial states in Sec.~\ref{sec:twoqubits}. In Sec.~\ref{sec:scaling} we discuss how ergotropy and coherence scales with the number qubits in the system for different classes of initial states. We conclude in Sec.~\ref{conclusions}.

\section{Preliminaries}\label{prelim}

\subsection{Ergotropy}\label{sec:ergo}

The term ergotropy, introduced and coined in the seminal paper by Allahverdyan \emph{et. al.}~\cite{EPL_Allahverdyan}, refers to the maximum amount of work that can be extracted from a quantum system through a cyclic unitary transformation of the initial state. Such a unitary transformation is generated by a time-dependent Hamiltonian $H_t=H+V(t)$, where $H$ is the self-Hamiltonian of the system that is used as a reference for the extracted energy and $V(t)$ is a time-dependent coupling to an external agent in which the work is deposited. The cyclic nature of the process is ensured by the conditions $V(t=0)=V(t=\tau)=0$ with $\tau$ being the duration of the work extraction process, which ensures that the system remains isolated in the beginning and at the end of the process. Now, let's assume that we are given a quantum state $\rho$ with its internal Hamiltonian $H$ such that they have the following spectral decomposition
\begin{align}
\rho&=\sum_jr_j\ket{r_j}\bra{r_j} & H&=\sum_i\varepsilon_i\ket{\varepsilon_i}\bra{\varepsilon_i},
\end{align}
where ordering of the eigenvalues for $\rho$ and $H$ is in decreasing, $r_1\geq r_2\geq\dots$, and increasing, $\varepsilon_1\leq\varepsilon_2\leq\dots$, order, respectively. Since unitary dynamics is considered, any decrease in the internal energy of the system at hand, with respect to its self-Hamiltonian $H$, will be extracted as work. Thus, in order to find the ergotropy one aims to minimize the internal energy of the final state 
\begin{equation}\label{eq:int_en}
\mathcal{W}=\text{tr}(\rho H)-\text{min}~\text{tr}(U\rho U^{\dag}H),
\end{equation}
where the minimization is performed over all possible unitaries. It has been shown in~\cite{EPL_Allahverdyan} that the final state $\rho_{\text{f}}\!=\!U\rho U^{\dag}$ that achieves this minimum has the form $\rho_{\text{f}}\!=\!\sum_jr_j\ket{\varepsilon_j}\bra{\varepsilon_j}$, i.e. a state that is diagonal in the energy eigenbasis, $[\rho, H]=0$, with its eigenvalues arranged in the decreasing order. A unitary operator which performs such a transformation is $U\!=\!\sum_j|\varepsilon_j\rangle\langle r_j|$ and when inserted in Eq.~\ref{eq:int_en} yields the following well-known expression for the ergotropy~\cite{EPL_Allahverdyan}
\begin{equation}\label{eq_ergo}
\mathcal{W}=\sum_{j,i}r_j\varepsilon_i(|\langle r_j|\varepsilon_i\rangle|^2-\delta_{ji}).
\end{equation}
It is also possible to find an explicit form for $V(t)$ by solving the Schr\"odinger Equation for the unitary operator $U$, however it does not yield a unique form~\cite{EPL_Allahverdyan}. A state that has its ergotropy, $\mathcal{W}$, equal to zero is called a passive state and any non-passive state is called an active state. Arrays of passive states can allow for work extraction from them through some collective process~\cite{NJP_Binder,PRL_Campiaoli,PRL_Ferraro,PRA_Le}, with thermal states being an exception. No combination of thermal states results in an active state, therefore they are called completely passive states~\cite{Pusz78}. 

\subsection{$l_1$ norm of coherence}

Following the pioneering work by Baumgratz \emph{et. al.}~\cite{PRL_Baumgratz} that set the ground rules that a proper coherence measure must satisfy, many different measures to characterize and quantify the coherence in a given quantum system have been introduced~\cite{RMP_coherence}. Among them, the $l_1$ norm is given by~\cite{PRL_Baumgratz}
\begin{equation}\label{eq_cl1}
C_{l_1}=\sum_{i\neq j}|\rho_{ij}|,
\end{equation}
which is simply the absolute sum of the off-diagonal elements of a given density matrix. 

In the following Sections, we will discuss how the steady-states of an open quantum system can possess non-zero ergotropy, i.e. be an active state, and its relation to the coherences in its energy eigenbasis. Such an active steady-state would be robust in preserving its ergotropy and work can be extracted from it after detaching it from the bath and going through the cyclic unitary process described above.

\section{Model}\label{model}

The model that we are going to discuss throughout this work is the many-particle generalization of the well-known quantum optical master equation~\cite{BreuerPetruccione,Lehmberg,Stephen,Gross1982}, which has been widely used in the literature~\cite{PRA_Latune,Damanet2016,PRA_Angsar,Bhattacharya2018}. We will assume that there are $N$ two-level systems (qubits from now on), which act like point-like dipoles and are assumed to have identical dipole moments, embedded in a thermal electromagnetic field environment. Depending on the spatial configuration of the system particles, they can be individually or collectively coupled to the environment, which we will discuss in a more detailed manner in what follows. The master equation governing the dynamics of such a system can be derived in usual Born, Markov approximations and assuming a weak coupling to the bath, which is given as follows (in units of $\hbar =1$)~\cite{PRA_Latune,Damanet2016,PRA_Angsar,Bhattacharya2018}
\begin{equation}\label{me}
\dot{\rho} = -i[(H_0+H_d), \rho]+\mathcal{D}_-(\rho)+\mathcal{D}_+(\rho)= \mathcal{L}(\rho).
\end{equation}
The first term on the right-hand side of the above equation accounts for the unitary evolution of the qubits where $H_0\!=\!\omega\sum_i^N\sigma_i^+\sigma_i^-$ is the free Hamiltonian of the qubits and $H_d\!=\!f_{ij}\sum_{i\neq j}\sigma_i^+\sigma_j^-$ is the dipole-dipole interaction between them with $f_{ij}$ being its strength and $\sigma_i^+\!=\!|e_i\rangle\langle g_i|$ and $\sigma_i^-\!=\!|g_i\rangle\langle e_i|$ are the raising and lowering operators for the $i$th qubit, respectively. The second and third terms describe the spontaneous and thermally induced emission (dissipation), and thermally induced absorption (incoherent driving) processes, respectively, whose explicit forms are as follows
\begin{equation}\label{dissipation}
  \mathcal{D}_-(\rho)=\sum\limits_{i,j=1}^N\gamma_{ij}(\bar{n}+1)(\sigma_j^-\rho\sigma_i^+-\frac{1}{2}\{\sigma_i^+\sigma_j^-, \rho\}),
\end{equation}
and
\begin{equation}\label{drive}
  \mathcal{D}_+(\rho)=\sum\limits_{i,j=1}^N\gamma_{ij}\bar{n}(\sigma_j^+\rho\sigma_i^--\frac{1}{2}\{\sigma_i^-\sigma_j^+, \rho\}),
\end{equation}
where $\bar{n}\!=\!(\exp(\beta\hbar\omega)-1)^{-1}$ is the mean number of thermal photons at the transition frequency of the qubits at an inverse temperature $\beta$.

The model above described by Eq. (\ref{me}), has two extreme limits. The first is when the qubits are far apart from each other compared to the wavelength of the photons in the environment. In this case, the qubits in the system behave as if they are individually coupled to the environment and mathematically this corresponds to $f_{ij}\approx 0$ and $\gamma_{ij}=\gamma_0\delta_{ij}$ with $\gamma_0=\omega^3d^2/3\pi\hbar\epsilon_0c^3$. Since the surrounding bath is a thermal one, naturally, all individual qubits thermalize with the temperature of the bath. The opposite extreme is the case when the qubits are closely packed such that the spatial separation between them is much smaller than the wavelength of the electromagnetic field in the environment. This is called the collective coupling limit with $f_{ij}\approx f$ and $\gamma_{ij}\approx \gamma_0$, and it has been shown that in this situation the total system evolves into a non-trivial steady-state that has coherences in the energy eigenbasis~\cite{PRA_Latune,PRA_Angsar}, even when the system is initiated in an incoherent state. The mechanism underlying the generation of these steady-state coherences (SSC) is the indistinguishability of the qubits to the bath in the collective coupling regime. Since it is not possible to know which qubit absorbed or emitted a photon and changed its state, overall system enters into a superposition state of such possible configurations. It is also important to note that in the collective coupling regime, the model does not admit a unique steady-state.

\section{Ergotropy of thermal coherences}\label{cohergo}

Throughout this work, we will be interested in the coherent steady-states that are generated by the dynamics dictated in the collective coupling limit. Intuitively, one would expect the steady-state of a system in contact with a thermal environment to be in a Gibbs state, which is a passive, even a completely passive state. However, due to the coherences generated and/or sustained as a result of the collective coupling of our system to the bath, we get a steady-state that do not have inverted populations, but still an active one such that it is possible to extract work from them through unitary cyclic processes~\cite{Korzekwa2016,PRX_Llobet,PRE_Thinga,francica,giacomo}. In other words, the very reason behind any finite ergotropy in the present setting is due to the coherences in the energy eigenbasis of the steady-state. It is also important to note that the coherences are generated as a result of the indistinguishability of two-level systems to the bath. Therefore, in order to obtain active steady-states one needs to have at least a pair of such subsystems; a single qubit would simply end up in a Gibbs (passive) state.

\subsection{Two qubits}\label{sec:twoqubits}

Since our model does not admit a unique steady-state, the steady-state that a given system reaches depends on its initial state. Recently, the analytical solution of Eq. (\ref{me}) in the collective coupling regime for an arbitrary initial state of a two qubit system is given in~\cite{PRA_Latune} as
\begin{align}\label{ss_twoqubit}
\rho_{ss}(\beta,c)&= (1-c)\ket{\psi_-}\bra{\psi_-} \\ \nonumber
& +cZ_+^{-1}\left(\beta\right)\left(e^{-2\omega\beta}\ket{\psi_1}\bra{\psi_1}+e^{-\omega\beta}\ket{\psi_+}\bra{\psi_+}+\ket{\psi_0}\bra{\psi_0}\right),
\end{align}
where $\ket{\psi_0}\!=\!\ket{00}$, $\ket{\psi_1}\!=\!\ket{11}$, $\ket{\psi_{\pm}}\!=\!\ket{01}\pm\ket{10}/\sqrt{2}$, $c\!=\!\bra{\psi_0}\rho_0\ket{\psi_0}+\bra{\psi_1}\rho_0\ket{\psi_1}+\bra{\psi_+}\rho_0\ket{\psi_+}$, and $Z_+\left(\beta\right)\!=\!1+e^{-\omega\beta}+e^{-2\omega\beta}$. We would like to note that in the present case $\ket{\psi_-}$ is stationary throughout the dynamics such that $\mathcal{L}(\ket{\psi_-}\bra{\psi_-})\!=\!0$. The thermodynamics of these steady-states have been extensively discussed in~\cite{PRA_Latune,PRR_Latune}. 
\begin{figure}[h]
{\bf (a)} \\
\includegraphics[width=0.8\columnwidth]{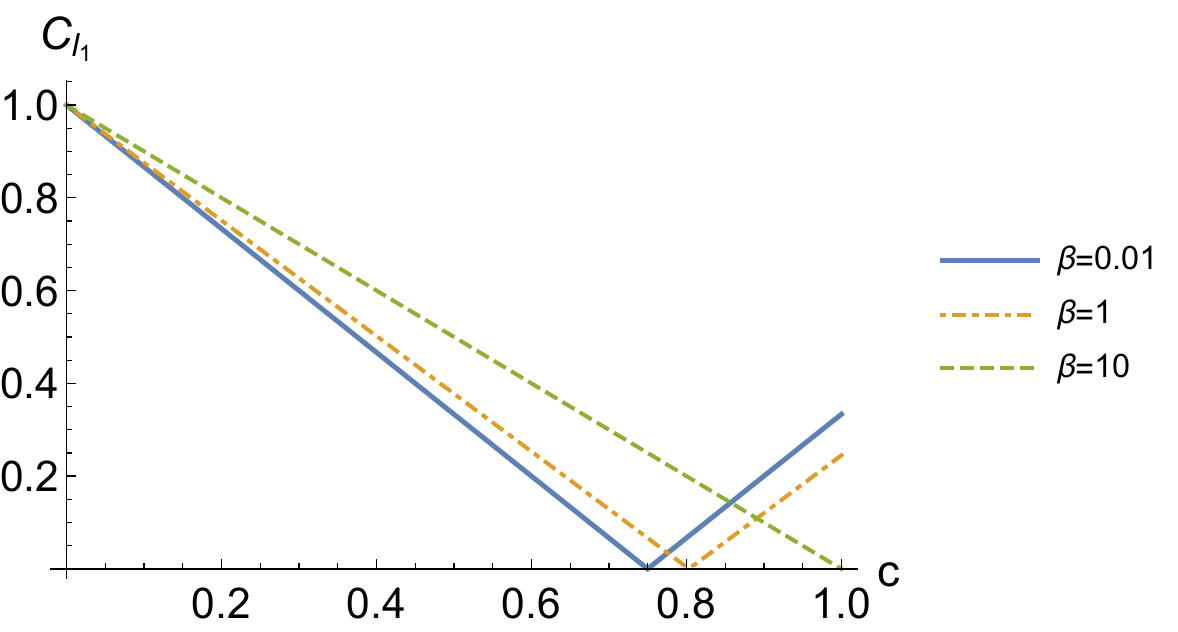}\\
{\bf (b)} \\
\includegraphics[width=0.8\columnwidth]{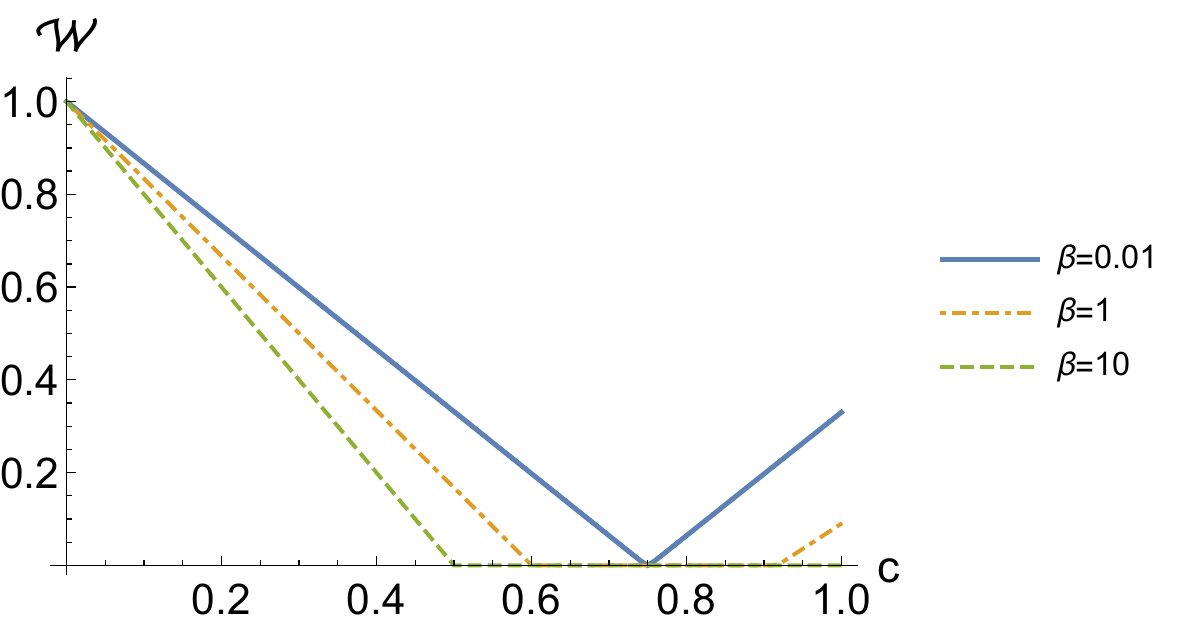}\\
\caption{{\bf (a)} Coherence as measured by $C_{l_1}$ and {\bf (b)} ergotropy calculated for different environment temperatures $\beta=0.01$ (solid), $\beta=1$ (dot-dashed) and $\beta=10$ (dashed) with $\omega=1$. }
\label{fig:twoqubits}
\end{figure}

We present our results on the coherence, as measured by the $l_1$ norm, $C_{l_1}$, and the ergotropy, $\mathcal{W}$, in Fig.~\ref{fig:twoqubits}. We observe that $C_{l_1}$ is always larger than $\mathcal{W}$ for all bath temperatures, with the exception of the high temperature limit, small $\beta$, where both quantities become equal to each other. Below, we will also analytically show that this is the case in the aforementioned limit. Note that the populations of the Eq. (\ref{ss_twoqubit}) are not inverted, therefore the ergotropy of these states is entirely due to the presence of the coherences.

In order to get a more comprehensive understanding of Fig.~\ref{fig:twoqubits}, we now would like to attempt to get analytical expressions for the $C_{l_1}$ and $\mathcal{W}$ for the present case of two qubits. A simple calculation shows that the amount of coherence in Eq. (\ref{ss_twoqubit}) as measured by the $l_1$ norm of coherence given in Eq. (\ref{eq_cl1}) is~\cite{PRA_Latune}
\begin{equation}\label{twoqubitcoh}
C_{l_1}=\left|c\frac{Z}{Z_+}-1\right|=\left|-1+c+\frac{c}{1+2\cosh(\beta\omega)}\right|,
\end{equation}
where $Z\left(\beta\right)\!=\!1+2e^{-\omega\beta}+e^{-2\omega\beta}$ is the partition function of the two qubit system. From the expression above, it is straightforward to determine the point at which $C_{l_1}$ vanishes as $c\!=\!Z_+/Z$ (cf. Fig~\ref{fig:twoqubits} {\bf (a)}). This implies that when the sum of the overlap of the initial state with $\ket{\psi_0}$, $\ket{\psi_+}$ and $\ket{\psi_1}$ is equal to its thermal value, i.e. the initial state is a thermal one, it is not possible for the dynamics to end up in a coherent steady-state. Note that as the temperature decreases the point of $C_{l_1}\!=\!0$ shifts towards $c\!=\!1$, since the thermal state at that temperature approaches to the ground state. Therefore, in order to reach a steady-state with coherences one needs to initiate the system in a state that has not equilibriated with the surrounding environment.

Next, we move on to the calculation of ergotropy defined in Eq. (\ref{eq_ergo}) for the steady-states we have at hand. It is straightforward to calculate the spectrum of the self-Hamiltonian of the qubits in our system, $H_0$, and in the ascending order it is given as $\{0,\omega,\omega,2\omega \}$ with their corresponding eigenvectors $\{ [0, 0, 0, 1]^T, [0, 0, 1, 0]^T, [0, 1, 0, 0]^T, [1, 0, 0, 0]^T \}$.~On the other hand, the ordering of the spectrum of $\rho_{ss}(\beta,c)$ is heavily dependent on the parameters characterizing the state. Below, we present the eigenvalues and corresponding eigenvectors and we will elaborate on the ordering problem later
\begin{align}\label{rhoeig}
&\{1 - c, \frac{ce^{2\beta\omega}}{1+e^{\beta\omega}+e^{2\beta\omega}}, \frac{ce^{\beta\omega}}{1+e^{\beta\omega}+e^{2\beta\omega}}, \frac{c}{1+e^{\beta\omega}+e^{2\beta\omega}}\} \\  \nonumber
&\left\{ \frac{1}{\sqrt{2}}[0, -1, 1, 0]^T, [0, 0, 0, 1]^T, \frac{1}{\sqrt{2}}[0, 1, 1, 0]^T, [1, 0, 0, 0]^T \right\}.
\end{align}
Recall that, in order to calculate the ergotropy we require that the eigenvalues of our state are in order of descending magnitude. For the tuple given above this condition is met when $c\leq 1/2$. Moreover for $c=1$, arranging the correct ordering is easy as the first eigenvalue is zero. Since the correct ordering is clearly determined for $c\!\leq\!1/2$ and $c\!=\!1$, we can obtain an analytical expression for the ergotropy in these regimes and it is given as
\begin{equation}
  \mathcal{W} =
  \begin{cases}
  \omega\left(1-\frac{c[1+3\cosh(\beta\omega)+\sinh(\beta\omega)]}{1+2\cosh(\beta\omega)}\right) & 0\leq c<1/2 \\
  \frac{\omega}{1+e^{-\omega\beta}+e^{-2\omega\beta}} & c=1, \\
  \end{cases}
\end{equation}
In the region $1/2\!<\!c\!<\!1$ the first eigenvalue gradually becomes smaller than the second, third and the fourth eigenvalue as $c$ increases up to $1$. We give the conditions that change the ordering of the eigenvalues in Appendix~\ref{appdxA}. Although it is involved to get an analytical expression for ergotropy for all $c$ at an arbitrary temperature due to the reasons stated above, we can obtain analytical expressions for low and high temperature limits. 

\emph{Low temperature limit - } In this regime we have $\beta\rightarrow\infty$, thus the eigenvalues given in Eq. (\ref{rhoeig}) reduces to $\{1-c,c,0,0 \}$. Clearly, the ordering of the eigenvalues change at the point $c\!=\!1/2$, yielding a step-wise behavior in ergotropy of the form
\begin{equation}
  \mathcal{W} =
  \begin{cases}
  \omega(1-2c) & 0\leq c<1/2 \\
  0 & 1/2\leq c\leq 1, \\
  \end{cases}
\end{equation}
which is consistent with the behavior we observe in Fig.~\ref{fig:twoqubits} {\bf (b)} for $\beta=10$. Note that $1-c$ measures the overlap that the initial state has with the anti-symmetric Bell state $\ket{\psi_-}$. Therefore, the result above implies that one needs to initiate the system in a state that has at least $1/2$ overlap with $\ket{\psi_-}$ to get a finite ergotropy at the steady-state in low temperatures.

\emph{High temperature limit -} This regime is characterized by $\beta\rightarrow 0$ which results in the eigenvalues given in Eq. (\ref{rhoeig}) to have the form $\{1-c,c/3,c/3,c/3 \}$. Clearly, we have a change in the ordering of the eigenvalues at the point $c=3/4$ that again gives rise to a step-wise behavior in the ergotropy given as
\begin{equation}\label{ergohighT}
  \mathcal{W} =
  \begin{cases}
  \omega\left(1-\frac{4c}{3}\right) & 0\leq c<3/4 \\
  0 & c=3/4 \\
  \omega\left(\frac{4c}{3}-1\right)   & 3/4< c\leq 1, \\
  \end{cases}
\end{equation}
which is again fully consistent with Fig.~\ref{fig:twoqubits} {\bf (b)} for $\beta=0.01$. At this point we would like to point an interesting connection. The $l_1$ norm of coherence for our two qubit system given in Eq. (\ref{twoqubitcoh}), reduces to $C_{l_1}=\left|-1+\frac{4c}{3}\right|$ in the present limit. This is exactly the same as the ergotropy per unit energy given in Eq. (\ref{ergohighT}), therefore, in the high temperature limit we have the following equality $\mathcal{W}/\omega=C_{l_1}$.

Related with the presented analysis we would like to discuss two relevant and natural questions. First, do quantum correlations give us any further insight about the origin of the finite ergotropy in these steady-states? Focusing on entanglement~\cite{Entanglement1,Entanglement2,EntanglementReview} and quantum discord~\cite{DiscordZurek,DiscordVedral,DiscordReview}, we can conclude that the answer is negative and we present their steady-state behavior as a function of $c$ in Appendix~\ref{appdxB} for comparison with Fig.~\ref{fig:twoqubits}. The entanglement content of the steady-states quickly vanishes with increasing temperature and remain finite only for small $c$ in the high-temperature limit, i.e. when the initial state has a sufficiently large overlap with $\ket{\psi_-}$. This is far from the very strict relationship we observe between ergotropy and $l_1$ norm in the high-temperature limit, that is $\mathcal{W}/\omega=C_{l_1}$, which leads us to conclude that the entanglement of the steady-state does not play a relevant role in the ergotropy content of the state, as compared to the coherence. On the other hand, the behavior of the quantum discord qualitatively follows the same trend with $C_{l_1}$ for different temperatures, albeit always smaller in magnitude, with equality attained when $C_{l_1}$ is zero, which is natural to expect since it is not possible to have a non-zero discord for a diagonal state, and at $c=1$. Similar to the case of entanglement, we do not see any further relevance between discord and ergotropy, as compared to the relevance we have observed with coherence and ergotropy.

Second question is related to the net change in the ergotropy of a given system. How does the ergotropy of an initial state at the beginning of the dynamics and its corresponding steady-state compare? Due to the open nature of our system, it is possible for some high internal energy initial states to dissipate some of their initially available energy during the time evolution. Therefore, depending on the initial state of the system, the considered dynamics does not always increase the ergotropy. For example, assume that we initiate our system such that both particles are in their excited state, $\ket{\psi_1}$, which corresponds to the steady-state with $c\!=\!1$. The initial ergotropy of such an initial state is $2\omega$ while its final ergotropy is definitely going to be less than that initial value, for all environment temperatures. However, another initial state that also correspond to the same steady-state with $c\!=\!1$ is when both qubits are initiated in their ground states, $\ket{\psi_0}$. Clearly, this initial state has zero ergotropy but it is brought to a state with non-zero ergotropy for a large range of temperatures. As compared to the unitary charging processes, this may seem a drawback of the presented method. Nevertheless, unitary charging protocols in general require high control over the system, energetically costly and assume perfect isolation of the battery from the surrounding environment, which is generally a challenging condition to meet for quantum systems. Our approach, on the other hand, require very little control both on the initial state preparation and ergotropy creation/storage steps, partially remove the necessity of perfect isolation of the subject system and it is energetically very cheap as compared to the unitary charging protocols since only a heat bath is needed. Moreover, due to the fact that the ergotropy is stored in the steady-state, it is robust and stable, which is an important issue in proposals for designing quantum batteries~\cite{PRE_Alan,arXiv_Rosa,PRR_Gherardini}.

Finally, we would like to briefly comment on the work extraction strategies for the active steady-states discussed in this section. As described in Sec.~\ref{sec:ergo}, by definition such a process is a unitary one so that any decrease in the internal energy of our system qubits can be considered as extracted work. Therefore, the system must be detached from the heat bath and processed afterwards. Furthermore, it is important to note that local states of the qubits are diagonal with non-inverted populations, meaning that they are passive. As a result, when processing these states to extract work, one must design a global process without discarding either one of the qubits. Such a global unitary can be constructed as $U\!=\!\sum_j\ket{\varepsilon_j}\bra{r_j}$~\cite{EPL_Allahverdyan} (also see Sec.~\ref{sec:ergo}), which simply takes the active steady-state in Eq.~\ref{ss_twoqubit} and diagonalizes it in the energy eigenbasis with decreasing populations, i.e. leaves it in a passive state. The explicit form of such a unitary can be easily determined from the eigenvectors of $\rho_{ss}(\beta,c)$ and $H_0$ presented in Eq.~\ref{rhoeig}. However, one must again pay attention to the ordering of the eigensystem of $\rho_{ss}(\beta,c)$, as we did in the calculation of the ergotropy, since it changes with the parameter $c$ that controls the initial state of the system.

\subsection{Scaling with the number of particles}\label{sec:scaling}

In this section, we would like to analyze how the ergotropy in the steady-state of our model scales with the number of particles and, in particular, how it compares with the scaling of $C_{l_1}$~\cite{PRA_Angsar}. However, as stated before, the model under consideration do not have a unique steady-state and we only have the analytical solution in the two qubit case for arbitrary initial states. As a result, in what follows we will analyze the cases with fixing the initial state to (i) ground initial states and (ii) random initial states.

However, before going into these cases, we would like to briefly comment on a different class of initial states. As we have already seen in the two qubit case, the dynamics under consideration does not generate any coherence at the steady-state for a thermal initial state, and thus ergotropy of is also equal to zero. This behavior also continues for larger number of qubits.

\subsubsection{Ground initial states}

We begin our discussion on the scaling with the case in which all qubits are initiated in their ground states. Such an initial state contains no coherence and clearly does not have inverted populations. Therefore, any amount of coherence and ergotropy is generated by the dynamics described by Eq. (\ref{me}). Remarkably, in this case it is possible to find an analytical expression for the steady-state of the system, as presented in~\cite{PRE_Angsar}, and it has the following block diagonal structure
\begin{equation}\label{rhoNss}
  \rho=
  \begin{pmatrix}
    D_{N} & 0 &\dots & 0 &0\\
    0 & D_{N-1} & \dots & 0 & 0\\
    \vdots & \vdots & \ddots & \vdots & \vdots \\
    0 & 0 & \dots &  D_1 & 0\\
    0 & 0 & \dots & 0 & D_0
  \end{pmatrix},
\end{equation}
where each block has the form has a size of $(p_k\times p_k)$ with $p_k=C(N, k)$. The explicit form of these blocks are given as $D_k=d_kU_k$ with $U_k$ is a matrix of ones and 
\begin{equation}
d_k=\frac{(1-r)r^k}{(1-r^{N+1})p_k},
\end{equation}
with $r=\bar{n}/\bar{n}+1$. The full derivation of this result can be found in~\cite{PRE_Angsar} (in particular see Appendix A of the mentioned reference).

The number of off-diagonal elements in a given block is $p_k^2-p_k$ with all of them being equal to $d_k$. Then, we can analytically calculate the $l_1$ norm by adding these up over all blocks which results in
\begin{equation}\label{cl1analytic}
C_{l_1}=\sum_{k=0}^Nd_kp_k(p_k-1)=\frac{(r-1) (r+1)^N}{r^{N+1}-1}-1.
\end{equation}
We can check a couple of relevant limits of the environment temperature for the above expression. In the low temperature limit we have $r\rightarrow 0$ and in turn $C_{l_1}\rightarrow 0$, which is expected since our initial state is actually the thermal state at zero temperature. In the opposite limit of high bath temperatures, that corresponds to $r\rightarrow 1$, we get $C_{l_1}=(2^N-N-1)/(N+1)$ which shows a $2^N$ scaling with the number of particles. Note that this limit gives us the highest amount of coherence that can be generated with the considered initial states under the considered dynamics. The physical mechanism behind this is as follows: Since all the qubits are initially in their ground state the only way to create coherence in this system is through the thermally induced absorption processes which happens at a rate of $\gamma_0 \bar{n}$. Therefore, increasing $\bar{n}$, which corresponds to increasing $\beta$, generates the highest amount coherence for this initial state~\cite{PRA_Angsar}.
\begin{figure}[t]

\includegraphics[width=0.8\columnwidth]{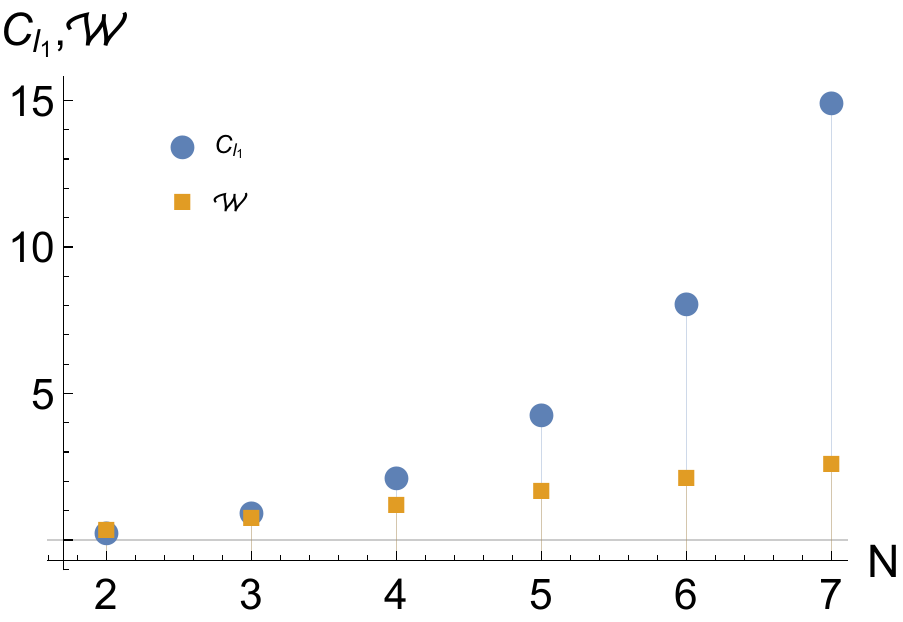}
\caption{Scaling of coherence (blue circles) and ergotropy (orange squares) as a function of the number of qubits initiated in their ground states in the high temperature limit, $\beta\rightarrow\infty$, and $\omega =1$.}
\label{scaling}
\end{figure}

Even though it is a bit more involved as compared to the calculation of $C_{l_1}$, it is also possible to obtain an analytical expression for the ergotropy for arbitrary number of particles which is given as 
\begin{equation}\label{ergoanalytic}
\mathcal{W}=\sum_{k=1}^{N-1}kp_{k+1}d_{k+1}\omega=\omega\left(N+\frac{N+r}{r^{N+1}-1}+\frac{r}{1-r}\right).
\end{equation}
We present the details of this calculation in Appendix~\ref{appdxC}. As expected, in the zero temperature limit, $r\rightarrow 0$, ergotropy goes to zero, since $C_{l_1}$ is also zero in this limit, which is actually our source of ergotropy in the present model. In the opposite end, as $r\rightarrow 1$, we obtain $\mathcal{W}=\omega[N(N-1)]/[2(N+1)]$. We can immediately observe the linear scaling in the ergotropy, in contrast to the $2^N$ scaling in the coherence at the high temperature limit. We present the scaling behavior in Fig.~\ref{scaling} at the high temperature limit where we get the highest amount of coherence and ergotropy up to $N=7$.

Eq. (\ref{ergoanalytic}) also allows us to comment on which blocks in the density matrix of our system contribute to the ergotropy together with the magnitude of their contribution.  Note that the ground state, $D_0$, and single excitation subspace, $D_1$, do not contribute to the ergotropy. While it is natural to expect $D_0$ not having any effect on the ergotropy, it is notable to see that coherences in the single excitation subspace also do not contribute to the ergotropy. All remaining blocks have a finite contribution which is proportional to their only non-zero eigenvalue $p_kd_k$. The magnitude of these eigenvalues decrease as go up in the block number, i.e. $p_kd_k>p_{k+1}d_{k+1}$ (see Appendix~\ref{appdxC}). This implies that coherences in the lower blocks have higher impact on ergotropy as compared to higher blocks, with the lowest two blocks being exceptions. 

At this point, a technical remark on the $N=2$ case is in order. Since the lowest two blocks do not affect the ergotropy, in the case of two-qubits initiated in their ground state, only contribution to the ergotropy comes from the $D_2$ block which is the population of the doubly excited state. While this may seem in contrast to our claims that the generated coherences are the actual source of the ergotropy, such a contribution is in fact enabled by the modification that the presence of coherences make in the spectrum of the density matrix. 

The reduced density matrices of each individual qubit of Eq. (\ref{rhoNss}) are also diagonal, similar to the previous section. However, when the number of qubits in the system is larger than two, it is possible to have coherences in the bipartite or larger sized reduced states. Therefore, although it is again not possible to get a finite ergotropy from individual qubits, it may be possible to extract work from combinations of the local states. Naturally, the amount one can get is smaller than that of the total, global state.

\subsubsection{Random initial states}

\begin{figure}[t]
{\bf (a)} \\
\includegraphics[width=0.8\columnwidth]{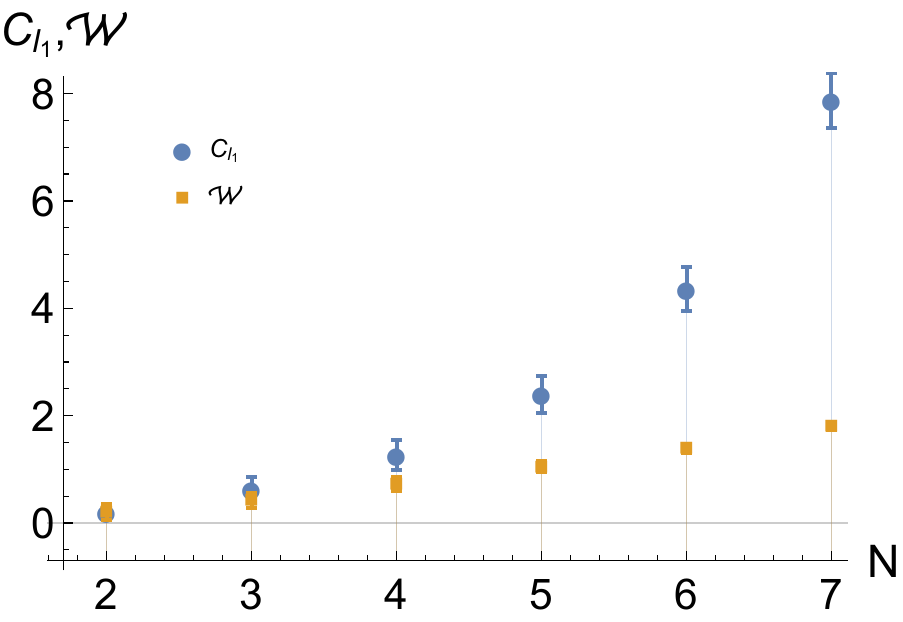}\\
{\bf (b)} \\
\includegraphics[width=0.8\columnwidth]{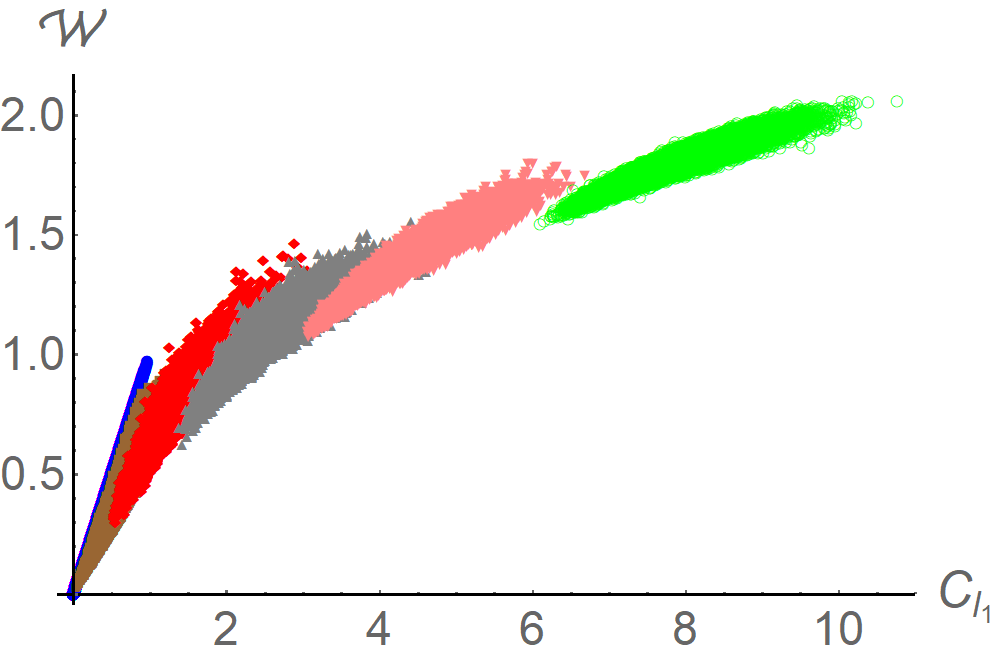}\\
\caption{{\bf (a)} Scaling of mean coherence (blue circles) and ergotropy (orange squares) as a function of the number of qubits initiated in $10^5$ random initial states in the high temperature limit, $\beta\rightarrow\infty$, and $\omega =1$. Error bars denote the standard deviation and if not visible they remain inside the data points. {\bf (b)} Ergotropy vs. coherence scatter plot for all $10^5$ random initial states for each two (blue), three (brown), four (red), five (gray), six (pink), seven (green) qubit systems. In gray scale stacks of points corresponding to $N=2$ to $N=7$ is displayed from dark to light, respectively.}
\label{fig:random}
\end{figure}

In this section, we present numerical scaling results for $10^5$ initial states for each system size up to $N=7$, to make a comparison with Fig. \ref{scaling}. We again investigate the high-temperature limit, since in this regime the initial states that have no coherences can end up in steady-states with higher amount of coherences, and therefore ergotropy, making this it more interesting and relevant. In Fig. \ref{fig:random} {\bf (a)} blue circles and orange squares mark the mean value of coherence and ergotropy,respectively, and error bars denote the standard deviation around these mean values. We observe that even though the mean coherence still grows more rapidly with the number of qubits as compared to the ergotropy, its growth rate is slower than it was for ground initial states. On the other hand, ergotropy shows a similar scaling behavior as it did for the ground initial states.

Fig. \ref{fig:random} {\bf (b)} shows a scatter plot of ergotropy vs. coherence again for all $10^5$ random initial states for each system size from $N=2$ to $N=7$. Each stack of points with a certain color corresponds to a different qubit number in the system and system size increases from left to right (see the figure caption for details). We observe that as the spread in the coherence gets larger, the spread in the ergotropy gets smaller and also seems to settle towards a smaller range of values as the number of particles increase in the system. This behavior can also be seen from the error bars representing the standard deviations in Fig. \ref{fig:random} {\bf (a)}. This suggests that as far as the ergotropy is concerned increasing the coherence in the system, which naturally can increase with the number of particles, can get less and less beneficial after a certain limit. As it would be a more challenging process to meet the close packing requirement of collective coupling with increasing number of particles, this result in fact shows us that one does not gain much by trying to achieve it.

\section{Conclusions}\label{conclusions}

We analyzed the steady-states of a system of qubits (two-level) systems in contact with a thermal environment in a collective manner. As a result of the collective coupling, when the system is composed of more than one qubit, these steady-states admit coherences in the energy eigenbasis due to their indistinguishability to the bath. We showed that solely due to the presence of such coherences, the steady-states generated by this open system dynamics yield a finite amount of ergotropy. In the case of two qubits we obtained an analytical expression for the ergotropy for a large number of initial states and showed that the amount of ergotropy per unit energy in the high bath temperature limit is equal to the $l_1$ norm of coherence. Further, we looked at the scaling behavior of both coherence and ergotropy with the number of qubits for two different classes of initial states which are all ground and random initial states. In the former case, we presented analytical expressions for both coherence and ergotropy for arbitrary number of qubits and observe a $2^N/N$ and $N$ scaling, respectively. For the latter case, we initiated our system in $10^5$ random initial states for all system sizes up to seven qubits and analyzed the mean values of coherence and ergotropy. We observed that coherence grows at a smaller rate while ergotropy shows a similar scaling with the number of particles, as compared to the ground initial state case. This suggests that presented approach is suitable for ergotropy generation and storage for a large class of initial states that do not require complex preparation stages.

We believe that it is interesting to see that it is possible to extract work due to the coherences created and/or sustained in the steady-state of a quantum system that is coupled to a dissipative heat bath, which usually have detrimental effects on the coherences, whereas vast majority of proposals in the charging process of quantum batteries employ unitary dynamics. Naturally, the presented method provides neither the highest amount of ergotropy nor the highest power as compared to the cases of closed charging processes. However it shows an example of an open quantum battery that is cheap in terms of control and resource requirements and in which the ergotropy is robust, since it is stored in the steady-state of the dynamics.

\begin{acknowledgments}
The author would like to thank Steve Campbell for various discussions throughout the development of this work. The author is supported by the BAGEP Award of the Science Academy and by The Research Fund of Bah\c{c}e\c{s}ehir University (BAUBAP) under project no: BAP.2019.02.03. 
\end{acknowledgments}

\bibliography{references}

\onecolumngrid
\appendix

\section{Ordering of eigenvalues between $1/2<c<1$}\label{appdxA}

In this region the first eigenvalue appearing on Eq. (\ref{rhoeig}) becomes gradually smaller than the second, third and fourth eigenvalue depending on $c$ and $\beta$. We present these conditions below 
\begin{equation}
1 - c<\frac{ce^{2\beta\omega}}{1+e^{\beta\omega}+e^{2\beta\omega}} \text{~~when~~}
  \begin{cases}
  \beta >\log \left(\frac{\sqrt{\frac{-7 c^2+10 c-3}{(2 c-1)^2}}(2c-1)-c+1}{2 (2 c-1)}\right) & \text{for~~$1/2< c\leq3/4$} \\
  \text{or} \\
  \beta\geq 0  & \text{for~~$3/4< c< 1$}, \\
  \end{cases}
\end{equation}
\begin{equation}
1 - c<\frac{ce^{\beta\omega}}{1+e^{\beta\omega}+e^{2\beta\omega}} \text{~~when~~} 0\leq\beta <\frac{\log \left(\frac{\sqrt{\frac{4 c-3}{(c-1)^2}} (c-1)-2 c+1}{2 (c-1)}\right)}{\omega }  \text{~~for~~$3/4< c< 1$}
\end{equation}
\begin{equation}
1 - c<\frac{c}{1+e^{\beta\omega}+e^{2\beta\omega}} \text{~~when~~} 0\leq\beta <\frac{\log \left(\frac{1}{2} \left(\sqrt{\frac{3-7 c}{c-1}}-1\right)\right)}{\omega }  \text{~~for~~$3/4< c< 1$}
\end{equation}
Note that the ordering among the second, third and fourth eigenvalues remains the same for all parameters. 

\section{Quantum correlations in the two-qubit steady-state}\label{appdxB}
The most common and easy way to quantify entanglement in two-qubit systems is to calculate concurrence~\cite{Entanglement1,Entanglement2,EntanglementReview}. It is given by the following expression
\begin{equation}
C(\rho)=\max \left\{ 0,\sqrt{\lambda_{1}}-\sqrt{\lambda_{2}}-\sqrt{\lambda_{3}}-\sqrt{\lambda_{4}},\right\},
\end{equation}
where $\{\lambda_{i}\}$ are the eigenvalues of the matrix $\rho \tilde{\rho}$ in decreasing order, with $\tilde{\rho}=(\sigma^{y}\otimes\sigma^{y})\rho^{*}(\sigma^{y}\otimes\sigma^{y})$.

Within the realm of measures that quantify quantum correlations which are more general than entanglement, quantum discord~\cite{DiscordZurek,DiscordVedral} clearly stand out in terms of its widespread usage in the literature and various applications~\cite{DiscordReview}. It is defined in terms of the discrepancy between two classically equivalent definitions of mutual information in the quantum regime
\begin{equation}
D^{\leftarrow}(\rho^{ab})=I(\rho^{ab})-J^{\leftarrow}(\rho^{ab}).
\end{equation}
Here, $I(\rho^{ab})=S(\rho^a)+S(\rho^b)-S(\rho^{ab})$ is the straightforward generalization of mutual information to quantum systems with $S(\rho)=-\text{tr}~\rho\text{log}\rho$ being the von Neumann entropy and $\rho^a$ and $\rho^b$ are the reduced states of the subsystems. On the other hand, $J^{\leftarrow}(\rho^{ab})=S(\rho^a)-\underset{\{\Pi_k^b\}}{\text{min}}\sum_kp_kS(\rho_k^a)$ where \{$\Pi_k^b$\} represent the set of all possible measurement operators that can be performed on subsystem $b$ and $\rho_k^a=(I\otimes \Pi_k^b)\rho^{ab}(I\otimes \Pi_k^b)/p_k$ are the post-measurement states of $a$ after obtaining the outcome $k$ with probability $p_k=\text{tr}(I^a\otimes\Pi_k^b \rho^{ab})$. $J^{\leftarrow}(\rho^{ab})$ can be interpreted as the maximum information gained about the subsystem $a$ by performing measurements on $b$ while creating the least disturbance on the overall quantum system and commonly called as the classical correlations. Quantum discord can be different from zero even when a quantum system is not entangled, but reduces to a measure of entanglement for pure states.

\begin{figure}[h]
{\bf (a)} \hskip0.4\columnwidth {\bf (b)}\\
\includegraphics[width=0.4\columnwidth]{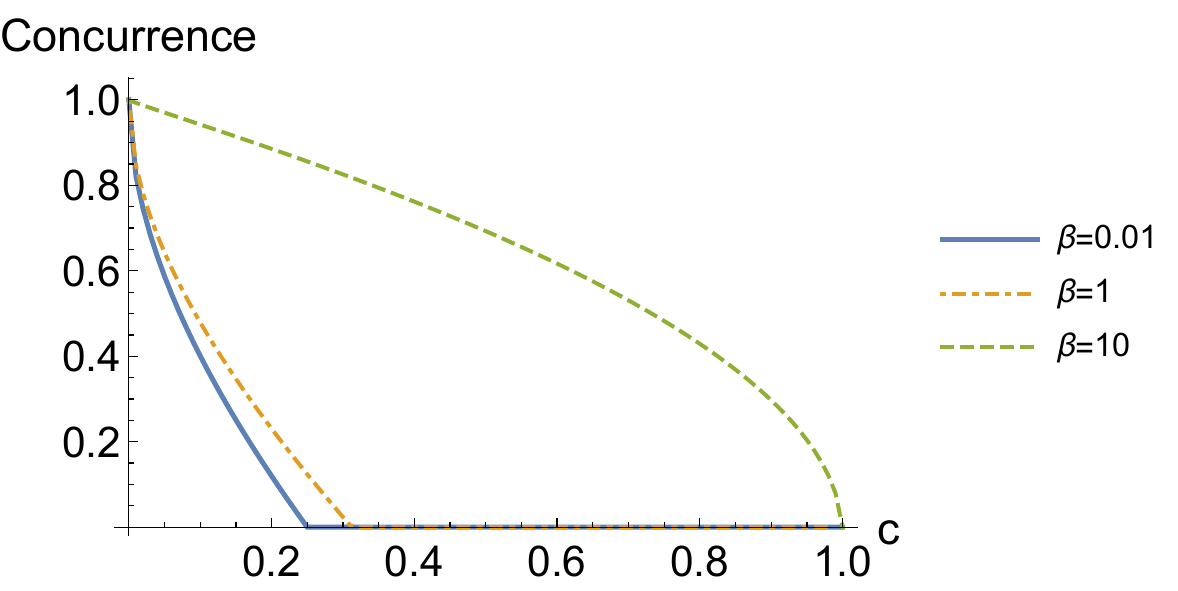}\includegraphics[width=0.4\columnwidth]{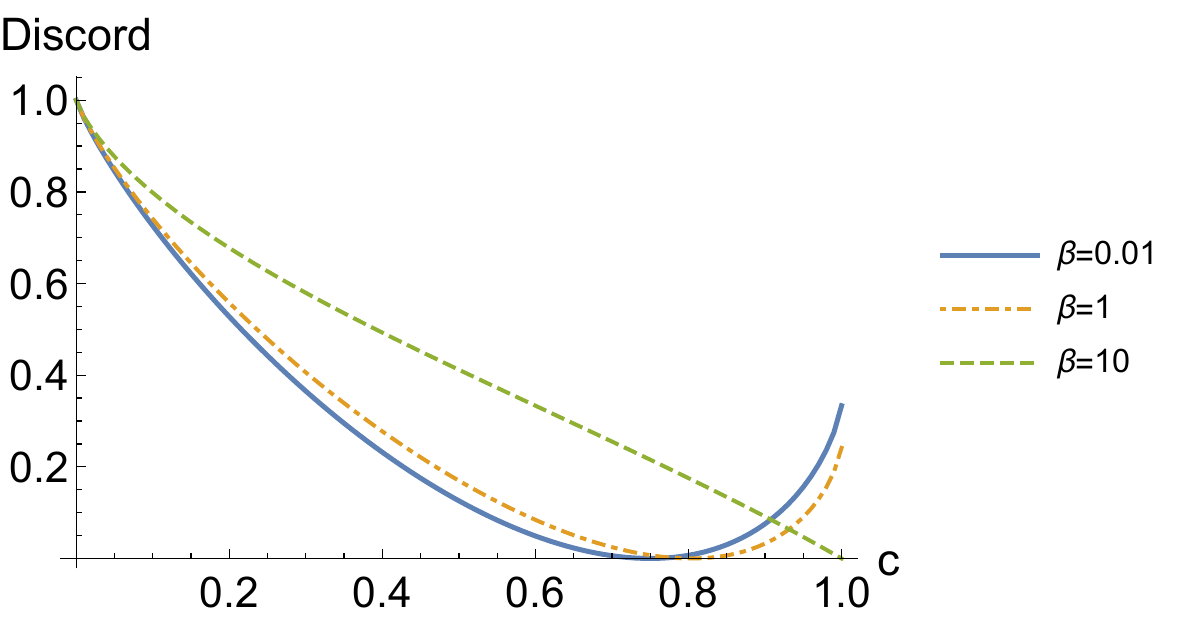}\\
\caption{Concurrence and quantum discord in the two-qubit steady-states of our model in the collective coupling limit for different temperatures of the bath. It is informative to compare these plots with that presented in Fig.~\ref{fig:twoqubits}}
\label{fig:correlations}
\end{figure}
In Fig~\ref{fig:correlations}, we present the entanglement and quantum discord as a function of our parameter that characterizes the initial state, therefore the steady-state of the system, $c$. We can see that the concurrence quickly becomes zero for a large range of $c$ as the temperature of the environment is increased which is in sharp contrast with the behavior of $l_1$ norm that remains non-zero except for a single value of $c$, that corresponds to thermal initial states, for all temperatures. On the other hand, quantum discord can be said to behave in a qualitatively similar fashion as the $l_1$ norm, but follow a smoother trend. Quantitatively, $D^{\leftarrow}\leq C_{l_1}$ and equality is attained when both of them are zero, i.e. when $c$ describes a thermal initial state, and when $c=1$. All in all, both concurrence and quantum discord do not show a stricter relation to ergotropy than $l_1$ norm presents.

\section{Calculation of ergotropy for ground initial states}\label{appdxC}

The ingredients we need for the calculation of ergotropy for arbitrary number of particles initiated in their ground-states are naturally the spectrum of the self-Hamiltonian of the particles and the spectrum of their steady-state. For $N$ particles, $H_0=\omega\sum_{i=1}^N\sigma_i^+\sigma_i^-$, and its spectrum in the increasing order given as 
\begin{align}
\varepsilon_1=& 0 & \ket{\varepsilon_1}=&[0,\dots,1]^T \\ \nonumber
\varepsilon_2=& \omega & \ket{\varepsilon_2}=&[\underbrace{0,\dots, 0}_{2^N-2} ,1, 0]^T \\ \nonumber
& \vdots & \vdots \\ \nonumber
\varepsilon_{p_1+1}=& \omega & \ket{\varepsilon_{p_1+1}}=&[\underbrace{0,\dots, 0}_{2^N-p_1-1}, 1,\underbrace{ 0, \dots, 0}_{p_1+1}]^T \\ \nonumber
\varepsilon_{p_1+2}=& 2\omega & \ket{\varepsilon_{p_1+2}}=&[\underbrace{0,\dots, 0}_{2^N-p_1}, 1,\underbrace{ 0, \dots, 0}_{p_1+2}]^T \\ \nonumber
& \vdots & \vdots \\ \nonumber
\varepsilon_{p_1+p_2+2}=& 3\omega & \ket{\varepsilon_{p_1+p_2+2}}=&[\underbrace{0,\dots, 0}_{2^N-p_1-p_2}, 1,\underbrace{ 0, \dots, 0}_{p_1+p_2+2}]^T \\ \nonumber
& \vdots & \vdots \\ \nonumber
\varepsilon_{2^N}=& N\omega & \ket{\varepsilon_{2^N}}=&[1,\dots,0]^T \nonumber
\end{align}

We now turn our attention to the spectrum of the steady-state for ground initial states given in Eq. (\ref{rhoNss}). The eigenvalues of a block diagonal matrix are the combination of the eigenvalues of each individual block. For $D_k$ we only have a single non-zero eigenvalue given as $p_kd_k$ with its corresponding eigenvector being $[1, 1,\dots, 1]^T/\sqrt{p_k}$ and rest of them equal to zero. As a result, in the spectrum of the steady-state we only have $N+1$ number of non-zero eigenvalues which can have a non-zero contribution to the ergotropy. The non-zero eigenvalues in the decreasing order with their corresponding eigenvectors is given as follows
\begin{align}
r_1=& p_0d_0 & \ket{r_1}=&[0,\dots,1]^T \\ \nonumber
r_2=& p_1d_1 & \ket{r_2}=&[\underbrace{0,\dots, 0}_{2^N-p_1-1} ,\underbrace{1,\dots, 1}_{p_1}, 0]^T/\sqrt{p_1} \\ \nonumber
r_3=& p_2d_2 & \ket{r_3}=&[\underbrace{0,\dots, 0}_{2^N-p_1-p_2-1} ,\underbrace{1,\dots, 1}_{p_2}, \underbrace{0,\dots, 0}_{p_1}, 0]^T/\sqrt{p_2} \\ \nonumber
& \vdots & \vdots \\ \nonumber
r_{N+1}=& p_Nd_N & \ket{r_{N+1}}=&[1,\dots,0]^T \\ \nonumber
\end{align}

We continue by first focusing on the diagonal terms in the ergotropy, i .e. $j=i$, and they are given as 
\begin{align}\label{ergodiag}
\sum_{j=1}^{N+1}r_j\varepsilon_j(|\langle r_j|\varepsilon_j\rangle|^2-1)= & 0+r_2\varepsilon_2\left( \frac{1}{p_1}-1 \right)-r_3\varepsilon_3-r_4\varepsilon_4-\dots-r_{N+1}\varepsilon_{N+1} \\ \nonumber
= & \frac{\omega p_1d_1}{p_1}-\omega p_1d_1-\omega p_2d_2-\dots-\omega p_Nd_N.
\end{align}
On the other hand, the off-diagonal terms, i.e. $j\neq i$, are
\begin{align}\label{ergooffdiag}
\sum_{j,i}r_j\varepsilon_i|\langle r_j|\varepsilon_i\rangle|^2 =& \frac{r_2\varepsilon_3}{p_1}+\frac{r_2\varepsilon_4}{p_1}+\dots+\frac{r_2\varepsilon_{p_1+1}}{p_1} &\Bigg\} & p_1-1~\text{terms} \\ \nonumber
\\ \nonumber
& +\frac{r_3\varepsilon_{p_1+2}}{p_2}+\dots+\frac{r_3\varepsilon_{p_1+p_2+1}}{p_2} & \Bigg\} & p_2~\text{terms} \\ \nonumber
\\ \nonumber
& +\frac{r_4\varepsilon_{p_1+p_2+2}}{p_3}+\dots+\frac{r_4\varepsilon_{p_1+p_2+p_3+1}}{p_3} & \Bigg\} & p_3~\text{terms} \\ \nonumber
\\ \nonumber
& \vdots & \\ \nonumber
\\ \nonumber
& +\frac{r_N\varepsilon_{p_1+p_2+\dots+p_{N-2}+2}}{p_{N-1}}+\dots+\frac{r_N\varepsilon_{p_1+p_2+\dots+p_{N-1}+1}}{p_{N-1}} & \Bigg\} & p_{N-1}~\text{terms} \\ \nonumber
\\ \nonumber
& +r_{N+1}\varepsilon_{2^N} 
\\ \nonumber
= & \left[p_1-1\right]\left[\frac{\omega p_1d_1}{p_1}\right]+p_2\left[\frac{2\omega p_2d_2}{p_2}\right]+\dots+p_{N-1}\left[\frac{(N-1)\omega p_{N-1}d_{N-1}}{p_{N-1}}\right]+N\omega p_Nd_N.
\end{align}
Combining Eq.~\ref{ergodiag} and Eq.~\ref{ergooffdiag} together we obtain the expression in Eq.~\ref{ergoanalytic} as follows
\begin{align}
\mathcal{W}= & \omega p_2d_2+2\omega p_3d_3+\dots+(N-1)\omega p_Nd_N \\ \nonumber
= & \sum_{k=1}^{N-1}k\omega p_{k+1}d_{k+1}. \\ \nonumber
\end{align}

\end{document}